\documentclass{optica-article}

\journal{opticajournal} 

\articletype{Research Article}

\usepackage{lineno}
\DeclareUnicodeCharacter{2062}{}

\begin{document}

\title{A three-dimensional acousto-optic deflector}

\author{Lewis R. B. Picard,\authormark{*} Manuel Endres}

\address{California Institute of Technology, Pasadena, California 91125, USA}

\email{\authormark{*}lpicard@caltech.edu} 


\begin{abstract*} 
Acousto-optic deflectors (AODs) are widely used across physics, microscopy, neuroscience, and laser engineering, providing fast, precise, and non-mechanical control of light. While conventional AODs naturally support multiplexing in one and two dimensions, no analogous device has existed for three-dimensional control, leaving a critical gap in rapid focus tuning and 3D beam shaping. Here we demonstrate a three-dimensional AOD system capable of multiplexed axial and lateral beam control with high speed and large dynamic range. We achieve this by combining a double-pass AOD with a diffraction grating in the Littrow configuration to realize a compact frequency-tunable lens with multiplexing capability. Our device enables axial scanning over more than twenty Rayleigh ranges with switching rates up to 100 kHz, while simultaneous multi-tone driving produces arbitrary multi-focal beam profiles. By integrating the axial module with lateral deflection, we generate reconfigurable 3D optical patterns. This approach establishes a broadly applicable platform for multiplexed 3D beam control, with potential applications from high-resolution microscopy and laser processing to scalable neutral-atom quantum technologies.
\end{abstract*}

\section{Introduction}

The ability to scan the focus of a laser beam is a key asset in areas including biological microscopy, advanced manufacturing, and quantum technology. Methods for rapid lateral beam scanning in 2D are widespread and mature, and include mechanical deflectors such as galvanometers and digital micromirror devices, reconfigurable spatial light modulators, and solid state devices such as electro-optic and acousto-optic deflectors (AODs) \cite{romerElectroopticAcoustoopticLaser2014}. The latter are routinely capable of achieving $\mu$s-scale movements. Unlike mechanical deflectors, AODs can be used for multiplexed scanning by driving the deflector with a multi-tone radio-frequency (RF) signal, which splits an input beam into multiple independently reconfigurable spots. Among other applications, in the context of quantum technologies, this allows for rapid rearrangement of atoms to densely filled arrays and the implementation of all-to-all connectivity in neutral atom quantum computers \cite{endresAtombyatomAssemblyDefectfree2016,bluvsteinLogicalQuantumProcessor2024}. Extending this rapid scanning capability to the axial direction to give full 3D reconfigurability requires some form of fast, focus-tunable lens, the design of which is an active area of research \cite{kangVariableOpticalElements2020}. There are various types of focus-tunable lenses based on mechanical deformation of the lens material, but these generally have response times on the millisecond scale and are incapable of multiplexing. One of the fastest known focus-tuning methods is acousto-optic lensing, which makes use of the cylindrical focusing produced by an AOD driven with a rapidly chirped RF signal \cite{kaplanAcoustoopticLensVery2001}. This method can yield focus-switching times of a few $\mu$s, but this comes at the cost of significant technical complexity, with focus tuning along a single axis requiring simultaneous application of precisely chirped signals to a series of four AODs in different orientations \cite{kirkbyCompactAcoustoopticLens2010}. Because of the coupling between axial focusing and lateral deflection in this method, it does not generally support multiplexing in 3D. Multiplexing in both lateral and axial dimensions can be achieved using spatial light modulators by imparting a programmable arbitrary phase to a wavefront, but even the fastest such modulators have limited update rates in the 1-10 kHz range \cite{linAIEnabledParallelAssembly2025,rochaFastLightefficientWavefront2024}.

\begin{figure}[htbp]
\centering\includegraphics[width=4.25in]{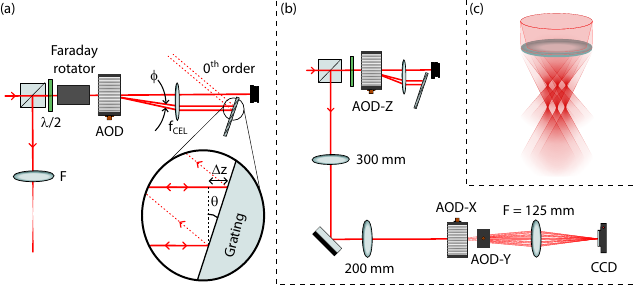}
\caption{ Illustrations of three-dimensional beam scanning system. (a) Optical setup of axial focus scanning module using a double-pass acousto-optic deflector (AOD) with a cat's eye lens (CEL) and a grating in the Littrow configuration. The grating has an angle $\theta$ relative to the incident beam, such that the first order diffracted beam is retroreflected along the incoming path. We indicate the optical path length difference $\Delta z$ between two beam deflection positions separated by an angle $\phi$, which results in a different defocus for each deflection angle. Because the AOD requires horizontal input polarization, and rotates the polarization of the diffracted light by 90$^\circ$, we use a Faraday rotator and a half-waveplate to separate the incoming and outgoing beams at a polarizing beam splitter. The beam is then re-focused to a spot with another achromatic lens with a focal length $F$. (b) Optical setup allowing for arbitrary scanning of a focal spot and multiplexed array generation in three dimensions using three AODs. The axial scanning module is imaged onto a pair of crossed AODs for x and y tuning using a 3:2 telescope in the 4-f configuration. (c) Illustration of a 3D array of optical tweezers created with our system, by sending superimposed beams with different defocus and input angles through a focusing lens.}
\label{fig:design}
\end{figure}

We describe herein a method for arbitrary rapid laser beam scanning and generation of complex optical patterns in three dimensions using AODs. The key innovation in our approach is the implementation of axial scanning using an AOD in a double-pass configuration with a cat's eye lens \cite{donleyDoublepassAcoustoopticModulator2005} and a diffraction grating in the Littrow configuration. The optical setup for axial scanning is illustrated in Fig. \ref{fig:design}(a). The cat's eye lens is placed one focal length away from the AOD, such that the variable angular deflection at the AOD crystal maps onto lateral displacement of the beam after the lens. The grating is placed approximately at the focus of the cat's eye lens, and its angle, $\theta$, relative to the incident beam is chosen such that the first-order diffracted beam is exactly counterpropagating with the incident beam (Littrow configuration). In this configuration the path length a beam travels between its forward and reverse passes through the cat's eye lens depends on its lateral position, which in turn is determined by the deflection angle at the AOD. As a result, tuning the deflection angle becomes equivalent to tuning the effective lens separation in the 1:1 telescope formed by the forward and reverse passes through the cat's eye lens. The output beam thus has a tunable amount of defocus. In the thin lens approximation, the whole system can be modeled as a lens with an effective focal length of
\begin{equation}
f=\frac{f_{CEL}}{2\mathrm{⁢tan}(\theta)\mathrm{⁢tan}(\phi)}, 
\end{equation}
where $f_{CEL}$ is the focal length of the cat's eye lens, $\theta$ is the Littrow angle of the diffraction grating, and $\phi$ is the angular displacement of the beam away from the angle which gives zero focal shift. By combining this axial scanning AOD module with two additional single-pass AODs for lateral scanning, as shown in Fig.\ref{fig:design}(b), we can achieve full 3D scanning and pattern generation, illustrated in Fig. \ref{fig:design}(c).

In a typical application, one would want to refocus the scanned beam using a downstream lens. In this case, the defocus introduced by the AOD lens translates to an axial shift of the beam waist away from the nominal focus of the refocusing lens. Crucially, because the angular deflections introduced by the AOD on the forward and reverse passes are equal and opposite to one another, the lateral position of the output beam does not change as its focus is tuned. Assuming that the AOD is one focal length behind a refocusing lens with focal length $F$, the shift of the waist is given by 
\begin{equation}
z_{s}=\frac{⁢F^2}{f}=\frac{2⁢F^2\mathrm{⁢tan}(\theta)\mathrm{⁢tan}(\phi)}{f_{CEL}}. 
\end{equation}
 We note that if the cat's eye and refocusing lenses do not form a perfect 4-f system, then the expression for the shift becomes more complex, with additional quadratic and higher-order dependence on $f$. Throughout this work we assume that the 4-f alignment is perfect. In order to write the shift in a way that is independent of the specific refocusing lens used, we can express it in units of the input beam waist at the AOD, $w_i$, and the Rayleigh range, $z_R = \frac{\pi w_0^2}{\lambda} = \frac{\lambda F^2}{\pi w_i^2}$, of the refocused beam. In units of $z_R$ the shift is
\begin{equation}
\frac{z_{s}}{z_R}=\frac{2\mathrm{tan}(\theta)\mathrm{tan}(\phi)\pi w_i^2}{\lambda f_{CEL}}. 
\end{equation}

From this expression, we can see that to achieve a large focus tuning range relative to the Rayleigh range one wants to use a beam with a large input waist, $w_i$, at the AOD. Subsequent magnification or demagnification of the beam will change both the spot size and the defocusing effect, leaving the ratio $\frac{z_{s}}{z_R}$ unchanged. There is, however, a tradeoff to using a large input beam, which is that the effective rise time of the AOD will increase, reducing the rate at which the focus can be tuned. In this work we used 813 nm laser light with an input beam waist of 2 mm and an AOD with an aperture of 7.5 mm (AAOptolectronic DTSX-400-810.920), which we chose to maximize the effective tuning range. This nevertheless allowed us to switch the focus at a rate of up to 100 kHz, as discussed further in Section 5. Throughout this work we used as our cat's eye lens a 1-inch diameter achromatic doublet (Thorlabs) with a focal length of $f_{CEL}=35$ mm, which we found provided a good compromise between ease of mounting and minimizing the cat's eye focal length. The diffraction grating is a ruled grating with a groove spacing of 1200 mm$^{-1}$ and a blaze angle of $32.7^\circ$ from Richardson Gratings. Because the AOD efficiency depends significantly on the polarization of the input beam, we placed a Faraday rotator before the AOD, which allowed us to align the polarization to maximize efficiency while still separating the input and output beams using a polarizing beam splitter cube. We measured the overall efficiency of the axial scanning module, which we define as the ratio of the optical power incident on the refocusing lens to the power immediately before the first pass through the AOD, to be 46\% at the center of the AOD bandwidth.

\section{Long-range focus tuning}

\begin{figure}[ht]
\centering\includegraphics[width=\textwidth]{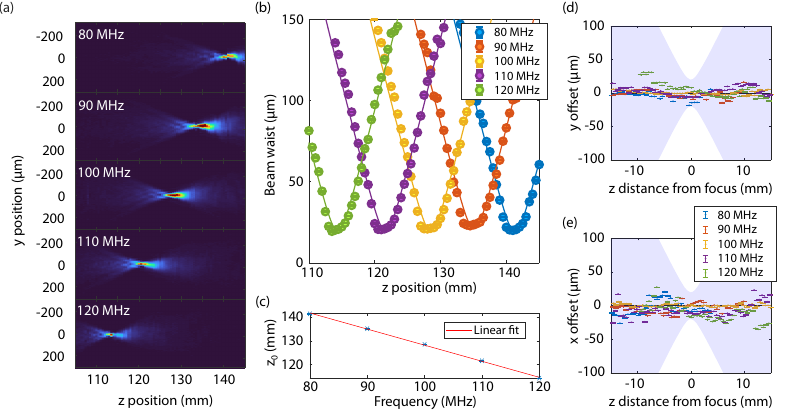}
\caption{ Characterization of long-range focus tuning module. (a) \textit{y-z} cut of the test beam profile for five different focus positions spanning the full bandwidth of the AOD used for focus tuning. All profiles are normalized to the maximum intensity at the center of the bandwidth. (b) Gaussian beam waist fits at various \textit{z}-positions for five different frequencies spanning the full bandwidth of the AOD used for focus tuning. The \textit{z} dependence of the waist is fit using a Gaussian profile with the Rayleigh range $z_R$, focus size $w_0$, and focus position as free parameters. (c) Dependence of focus position along \textit{z} on AOD frequency, showing the linearity of focus tuning with AOD deflection angle, as predicted by Eq. 1. (d) Displacement along \textit{z}, perpendicular to the AOD axis, of the beams generated by each RF tone, as a function of distance from the focus of that beam. The shaded area is a guide to the eye showing the beam waist as a function of \textit{z}, using the fit paraemeters from the central (100 MHz) spot. (e) Displacement along \textit{z}, parallel to the AOD axis, as a function of distance from the focus. }
\label{fig:waistFits}
\end{figure}

We first demonstrated the performance of the axial scanning module by tuning the focus position of a beam over a large range using a wide-bandwidth AOD.  We chose not to use the Faraday rotator in testing the tuning range of the device, as the particular rotator we used had a smaller aperture (5 mm) than the AOD itself, which would limit our input beam size. This reduces the overall power efficiency of the device, but otherwise leaves its behavior unchanged. For testing, we used an achromatic refocusing lens with a focal length of $F=125$ mm.

In Fig. \ref{fig:waistFits}(a), we show \textit{y-z} cuts of the beam profile for five different focus positions, each corresponding to a different AOD frequency. The AOD has a center frequency of 100 MHz and a nominal bandwidth of 36 MHz, corresponding to a 44 mrad scan angle, which we slightly exceeded with test frequencies which span 80 to 120 MHz. The mean observed efficiency of the device at the extrema of the scan range is approximately 28\% of the peak efficiency at the center frequency. The data were collected by scanning the \textit{z}-position of a CCD camera across a 50 mm range and taking images at 0.5 mm intervals, allowing us to reconstruct a three-dimensional intensity profile of the beam. We characterized the beams at different focal lengths in more detail by fitting a Gaussian profile to them at each \textit{z}-position. We then performed a variance-weighted fit of the dependence of the beam waist on \textit{z} to an ideal Gaussian beam model, $w(z) = w_0\sqrt{1+\left(\frac{(z-z_0)}{z_R}\right)^2}$, with the focus waist, $w_0$, Rayleigh range, $z_R$, and focus position, $z_0$, as free parameters. These fits are shown in Fig. \ref{fig:waistFits}(b). Across all of the beams we measure a mean Rayleigh range of 1.21(3) mm and a mean focus waist 19.6(3) $\mu$m. We do not measure any significant variation of $w_0$ across this tuning range, although all of the beams suffer from some astigmatism, which causes slight deviation from the ideal symmetric Gaussian beam model used for fitting. The separation between the focus positions for the 80 MHz and 120 MHz setpoints of the AOD is 26.9(2) mm. Using Eq 2, we see that this is equivalent to a minimum achievable focal length (maximum dioptric power) for the AOD lens of $f = 580(4)$ mm. Using the mean of the fitted $z_R$ values of 1.21(3) mm, we find the tuning range in units of the Rayleigh range to be $\frac{z_s}{z_R} = 22.3(5)$. In Fig. \ref{fig:waistFits}(c) we perform a linear fit of the focus positions as a function of AOD frequency, which is an excellent approximation to the $\textrm{tan}(\phi)$ dependence predicted by Eq. 1 for the typical deflection angles of the AOD. We observe a very small deviation from linearity at the edges of the bandwidth, which we attribute to a possible slight misalignment of the 4-f system, which introduces a quadratic dependence on $\phi$. We also measure the displacement of the beams from the focus for different focus-tuning positions, which we plot in Fig. \ref{fig:waistFits}(d,e). To remove any common-mode displacements resulting from the alignment of the beam axis relative to the camera translation stage, we define the origin for all images taken in a particular \textit{z}-plane to be a geometric centroid of the intensity distribution for the central, 100 MHz, focus position. We find that the focii for all of the different tones overlap along the \textit{x} and \textit{y} directions to within one focus beam waist. The standard deviations of the measured focus displacements from the origin, taken across all tones and \textit{z}-positions, are 11.7 $\mu$m along \textit{x} and 8.5 $\mu$m along \textit{y}. 

We note that this axial scanning module design could be used to achieve significantly larger scan ranges by reducing the focal length of the cat's eye lens. The lens, grating, and AOD in our test setup are mounted on an optical breadboard with standard 1-inch posts and optomechanics, which constrains the minimum achievable separation of these components. If one were instead to use smaller optics, or develop a custom mount to hold the AOD crystal, lens, and grating, it should be straightforward to reduce $f_{CEL}$ by a factor of four or more. This would result in a proportional increase in the scan range to approximately $100z_R$.

\section{Engineering multi-focal intensity profiles}

\begin{figure}[htbp]
\centering\includegraphics[width=3in]{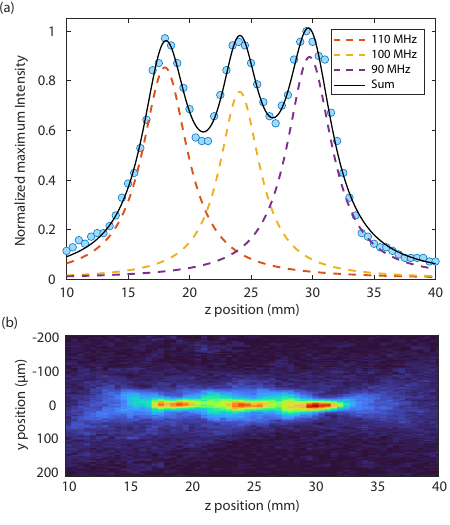}
\caption{Characterization of axially multiplexed beam. (a) Maximum intensity at each \textit{z}-position of camera as it is scanned along the axis of an axially multiplexed beam. Fit is performed to a sum of Gaussian beam profiles. (b) \textit{y-z} cut of the intensity profile of the multiplexed beam.}
\label{fig:multitone}
\end{figure}

A key feature of the axial scanning method presented here, which distinguishes it from other focus-tuning devices, is the ability to generate axial intensity profiles with multiple focii. Such axially multiplexed beams may have a wide range of applications, including the creation of 3D arrays of optical tweezers for atom trapping \cite{barredoSyntheticThreedimensionalAtomic2018,kusanoPlaneselectiveManipulationsNuclear2025,linAIEnabledParallelAssembly2025,schlosserScalableMultilayerArchitecture2023}, 3D imaging of biological samples \cite{maurerDepthFieldMultiplexing2010,duanAxialBeamScanning2017}, and industrial manufacturing \cite{duMultifocalLaserProcessing2022,volppImpactMultifocusBeam2019}. An axially multiplexed beam can be generated by driving the AOD simultaneously with multiple RF tones, similarly to how AODs are used to split a single beam into a line of resolvable spots to generate a 1D array of optical tweezers \cite{endresAtombyatomAssemblyDefectfree2016}. In the case of the axial scanning AOD, the different AOD frequencies will produce a line of spots on the grating and, after the second pass through the AOD, multiple copropagating beams with different amounts of defocus. The second pass through the AOD will also introduce undesirable cross-terms, since each of the beams on the second pass through the AOD will again be diffracted by all the RF tones. However, only those beams for which the forward-pass diffraction angle $\phi_f$ is equal to the reverse-pass diffraction angle $\phi_r$ will emerge copragating with the input beam. All other components will emerge at an angle $\phi_f - \phi_r$ relative to the input beam, allowing them to be spatially filtered out using a downstream aperture. The cross-terms will, however, still lead to a reduction in the power efficiency of the device by a factor of $1/N$, where $N$ is the number of RF tones.

In Fig. \ref{fig:multitone}, we demonstrate this capability by driving the AOD simultaneously with three RF tones at 90, 100, and 110 MHz. We refocused the beam using the same $F = 125$ mm lens as in Section 2, and imaged it at a range of \textit{z}-positions about the focus using a camera on a translation stage. We empirically optimized the amplitudes of the RF tones to approximately equalize the intensity at each focus. In Fig. \ref{fig:multitone}(a) we plot the peak intensity in each image as a function of \textit{z}-position. We fit the overall intensity profile to a sum of three Gaussian beam profiles, with the focus position, amplitudes and Rayleigh ranges of each beam as free parameters. In Fig. \ref{fig:multitone}(b), we show a \textit{y-z} cut of the multifocal intensity profile. We note that, because the AOD also shifts the frequency of each diffracted beam by the acoustic drive frequency, the three beams which are summed to create the multifocal intensity profile have slightly different optical frequencies. This means that there will be some time-dependent optical interference between the beams, which would manifest as rapid oscillation of the intensity in regions with nonzero overlap between different tones. Because the beams are generated with 90, 100, and 110 MHz driving frequencies, we would expect beating at 10 and 20 MHz, much faster than the frame rate of the camera used to measure the intensity profile. We anticipate that in most use cases such fast beating will not significantly affect performance. For instance, in the case of neutral atom trapping, the MHz-scale frequency separations of the RF tones are approximately one-to-two orders of magnitude larger than typical atom trapping frequencies, meaning that modulation of traps at these frequencies will have a minimal effect on atom temperature \cite{wangReductionLaserIntensity2020}.

\section{Full 3D focus pattern generation}

\begin{figure}[htbp]
\centering\includegraphics[width=4.25in]{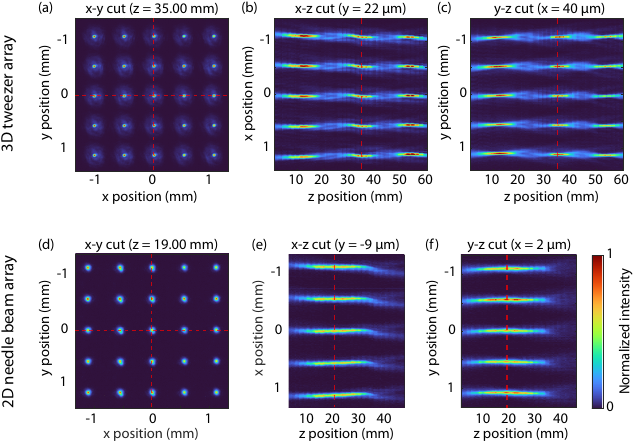}
\caption{Profiles of 3D intensity patterns generated using three AODs. (a-c) Images showing \textit{x-y}, \textit{x-z} and \textit{y-z} cuts of a 5-by-5-by-3 array of focused optical-tweezer-like spots generated using the setup. Dashed lines in each image indicate the planes along which the orthogonal cuts are displayed. (d-f) \textit{x-y}, \textit{x-z} and \textit{y-z} cuts of a 5-by-5 array of needle beams.}
\label{fig:3DArray}
\end{figure}

By combining the axial scanning module with additional AODs for lateral deflection, we can generate reconfigurable focus patterns in three dimensions. Fig. \ref{fig:design}(b) shows the optical setup we used to achieve this, with a pair of crossed AODs placed after the axial scanning module in a 4-f configuration. We used the same AAOptolectronic DTSX-400 AODs for lateral scanning as for axial. In Fig.\ref{fig:3DArray}(a-c), we show cuts along 3 orthogonal planes of a 5-by-5-by-3 array of focused spots. We created this array by driving the axial scanning AOD with three RF tones and each of the lateral scanning AODs with five tones. The multitone RF signals were generated with an arbitrary waveform generator from Spectrum Instrumentation (M4i.6622). If applied to optical tweezer generation for ultracold atom trapping, this setup would allow us to generate a reconfigurable 3D array of tweezer-trapped atoms. We note that there is a slight compression (expansion) of the lateral array spacing along \textit{x} (\textit{y}) as a function of \textit{z}-position, which results from the fact that we place the \textit{x} (\textit{y}) AOD immediately after (before) the true image plane of the 4-f system. This slight distortion of the array could be mitigated by adding an additional 4-f optical relay between the two lateral scanning AODs, which would allow each of them to be placed exactly in the image plane.

In Fig.\ref{fig:3DArray}(d-f) we used the flexibility of the three-dimensional scanning configuration to generate a 5-by-5 array of elongated focus \textit{needle beams}. Each of these beams was formed by driving the axial scanning AOD with five closely-spaced RF tones, creating multiple unresolved focal spots along z. This leads to a total intensity profile for each beam with an elongated needle-like focus. Needle beams generated with diffractive optical elements have been applied to the microscopy of biological samples to increase the depth of field of images \cite{caoOpticalresolutionPhotoacousticMicroscopy2023}. Our method would allow such needle beams to be rapidly reconfigured to yield different depths of field, as well as providing the ability to rapidly scan multiple needle beams in parallel using the lateral AODs. Such reconfigurable needle beams would also have immediate applications in precision laser machining, where increased depth of field can improve the efficiency of machining in cases where the surface height of a material varies during machining \cite{chenUltrafastZscanningHighefficiency2018}.

\section{Rapid focus switching}

\begin{figure}[htbp]
\centering\includegraphics[width=4.25in]{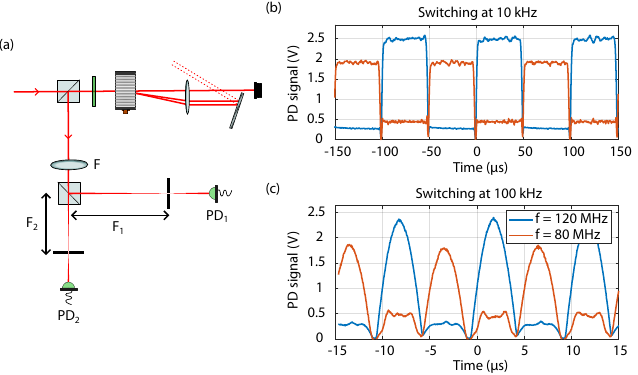}
\caption{Characterization of switching rate of the acousto-optic lens. (a) Optical setup used for testing rapid focus switching. With a focusing lens of $F = 250$ mm, the separation between the focii $F_2-F_1$ is approximately 90 mm. The irises in each path are aligned so as to maximize transmission of one tone and minimize background transmission of the other. (b) Signals of each photodiode (PD) when switching between two focus positions at 10 kHz. (c) Switching between two focus positions at 100 kHz.}
\label{fig:switching}
\end{figure}

In addition to its multiplexing capabilities, a key reason to use an AOD for axial scanning, rather than a mechanical or liquid crystal varifocal lens, is the possibility of very fast focus switching \cite{kangVariableOpticalElements2020}. The tuning bandwidth of the AOD lens is limited by the finite rise time of the AOD, which is determined by the time it takes for an acoustic wave in the crystal to traverse the width of the beam. In the demonstration presented here, we optimized for the largest practical tuning range. As such, we used a large input beam at the AOD, and selected a commercially available shear-mode AOD with a large angular bandwidth, which is due in part to the lower acoustic velocity (650 m/s) in this mode. This reduces the speed of focus tuning due to the finite time it takes an acoustic wave to traverse the full extent of the beam inside the crystal. The nominal rise time for the AOD we use, defined as the time for the power measured at one focus position to go from 10\% to 90\% of its maximum value, is 1 $\mu$s/mm beam diameter. For our 2 mm input beam we measured a rise time of 2 $\mu$s, consistent with this specification. 

We tested rapid switching of the focal position using the setup illustrated in Fig. \ref{fig:switching}(a). After a refocusing lens, we split the beam into two arms with equal powers using a polarizing beam splitter, and in each arm we placed an iris close to the focal position for AOD frequencies of 80 and 120 MHz, respectively. This allowed us to block a large fraction of the power for one focus position in each arm, while transmitting most of the power for the other focus position. We placed photodiodes behind the irises to measure the optical power. For this measurement we used a refocusing lens with a focal length of $F = 250$ mm to allow for easier discrimination between the focus positions. In Fig. \ref{fig:switching}(b,c) we show the measured signal at each photodiode during frequency modulation of the AOD drive RF signal with square waves at 10 and 100 kHz, respectively. We find that switching remains viable up to 100 kHz, albeit with a reduced average power due to the finite switching time. We measure an average reduction in the mean square value of the photodiode signals by a factor of 2.1, or 3.2 dB, when we increase the switching rate from 10 to 100 kHz.

If the design of the axial scanning module were optimized for speed instead of tuning range, we believe that response times of less than 1 $\mu$s could straightforwardly be achieved by using a smaller beam and/or selecting an AOD crystal with a higher acoustic velocity. Longitudinal mode $\mathrm{TeO}_2$ AODs typically have an acoustic velocity of 4200 m/s, so using such an AOD in this setup would allow for switching rates of over 600 kHz, albeit at the expense of a reduced scan range. An additional practical limitation on the rate at which the focus can be continuously tuned is the cylindrical focusing effect of the AOD crystal, which was itself used as the basis for previous acousto-optic lens designs in Refs. \cite{kaplanAcoustoopticLensVery2001,kirkbyCompactAcoustoopticLens2010}. The amount of cylindrical defocus is determined by the chirp rate of the AOD driving frequency, with the effective focal length for one pass of the AOD being $F_{cyl} = \frac{v^2}{\lambda\frac{df}{dt}}$, where $v$ is the acoustic velocity in the crystal and $f$ is the driving frequency \cite{kaplanAcoustoopticLensVery2001}. For the full AOD lens setup considered in Fig.\ref{fig:design}(a), this would result in a shift at the focus position of $\frac{z_{cyl}}{z_R}=\frac{2\pi w_i^2}{v^2}\frac{df}{dt}$. In cases where the beam shape must remain relatively constant while the focus position is being changed, for instance in coherent transport of neutral atoms in optical tweezers, this cylindrical defocus will become the dominant limit on the focusing rate. If one requires that the cylindrical focus shift remain always below $1z_R$, this would impose a speed limit of 17 kHz/$\mu$s for the AOD parameters considered in this setup. Traversing the full focus-tuning range shown in Fig. \ref{fig:waistFits} at this rate would take 235 $\mu$s, which remains well within the range of useful transport times for atoms in tweezers. For applications where the focus needs to be rapidly switched but the beam shape during the switching time does not matter, much faster tuning can be achieved by discretely switching the driving frequency, as we do for the data in Fig. \ref{fig:switching}. In applications where the focus is continuously scanned at a constant rate, such as in a raster scanning apparatus for laser manufacturing, the amount of cylindrical defocus will be constant, allowing it to be passively corrected with an additional cylindrical lens if needed.

\section{Conclusion and outlook}

We have demonstrated that a series of three AODs can be used to rapidly scan the position of a beam in three dimensions, and that by multiplexing the AOD control signals we can generate complex, reconfigurable optical patterns in 3D. The axial scanning component of our method combines two well-known optical techniques, an AOD in a double-pass configuration and a diffraction grating in the Littrow configuration, to produce a focus-tunable lens controlled by the frequency of a single driving RF tone. We measured the tuning range of axial scanning by refocusing the light onto a camera and mapping out the focus position. We found that we can achieve a focal length range of $f = \infty$ to $580(4)$ mm for the AOD effective lens, which, after refocusing, allows us to tune the position of the focus by 22.3(5) times the Rayleigh range. By driving the AOD simultaneously with multiple RF tones, we demonstrated that this setup can be used to engineer complex axial beam profiles. By combining the axial scanning module with additional AODs for lateral scanning, we demonstrated the creation of 3D arrays of resolvable spots and 2D arrays of elongated-focus needle beams. Finally, we showed that the lens can rapidly switch between the maximum and minimum focus positions at a rate of up to 100 kHz, with a rise time of just 2 $\mu$s. 

We expect that this device could be applied in neutral atom quantum technologies to enable trapping and fast rearrangement of single atoms in three dimensions. Although some instances of rearrangement of atom arrays in 3D have been demonstrated, the rate of motion in the axial direction has been limited by the slow response times of the varifocal devices used \cite{barredoSyntheticThreedimensionalAtomic2018,linAIEnabledParallelAssembly2025}. Using the 3D array generation and rapid scanning capabilities demonstrated here, one could generate large 2D tweezer arrays that can be rapidly translated in the axial direction, allowing for efficient transport of many atoms between planes of an auxiliary static 3D array generated by, for example, an SLM, optical lattice, or microlens array \cite{kusanoPlaneselectiveManipulationsNuclear2025,schlosserScalableMultilayerArchitecture2023}. Rapid axial focus tuning may also find applications in long-distance optical transport of atoms and molecules between different regions of a vacuum chamber, a task that is currently achieved using slow mechanical varifocal lenses \cite{leonardOpticalTransportManipulation2014,unnikrishnanLongDistanceOptical2021,baoFastOpticalTransport2022}.

\section{Back matter}

\begin{backmatter}
\bmsection{Funding} We acknowledge support from the NSF QLCI program (2016245).

\bmsection{Acknowledgment}
We thank Elie Bataille for fruitful conversations during the preparation of this manuscript. The authors declare no conflicts of interest.

\bmsection{Data availability} Data underlying the results presented in this paper may be obtained from the authors upon reasonable request.

\end{backmatter}

\bibliography{Bibliography}

\end{document}